# Two-photon interference of temporally separated photons

Heonoh Kim[1], Sang Min Lee[1,2], and Han Seb Moon[1]

[1]Department of Physics, Pusan National University, Geumjeong-Gu, Busan 609-735, Korea

[2]Currently with Korea Research Institute of Standards and Science, Daejeon 305-340, Korea

## Abstract

We present experimental demonstrations of two-photon interference involving temporally separated photons within two types of interferometers: a Mach-Zehnder interferometer and a polarization-based Michelson interferometer. The two-photon states are probabilistically prepared in a symmetrically superposed state within the two interferometer arms by introducing a large time delay between two input photons; this state is composed of two temporally separated photons, which are in two different or the same spatial modes. We then observe two-photon interference fringes involving both the Hong-Ou-Mandel interference effect and the interference of path-entangled two-photon states simultaneously in a single interferometric setup. The observed two-photon interference fringes provide simultaneous observation of the interferometric properties of the single-photon and two-photon wavepackets. The observations can also facilitate a more comprehensive understanding of the origins of the interference phenomena arising from spatially bunched/anti-bunched two-photon states comprised of two temporally separated photons within the interferometer arms.

The Hong-Ou-Mandel (HOM) interference effect[1] and the interference of path-entangled two-photon states[2], i.e., the so-called N00N state[3], have played an important role in fundamental investigations of quantum mechanics and the exploration of quantum information technology[4,5]. Since the late 1980s, various kinds of two-photon interference experiments have been performed in order to distinguish quantum mechanical treatment of optical interference phenomena from conventional classical optics[6,7]. These experiments have successfully shown that the interferences of correlated photons cannot be explained by any classical wave theory; instead, they should be viewed as interference between superposed probability amplitudes. The coherent superposition of states and the interference between probability amplitudes for indistinguishable processes in the total detection process have a crucial role in quantum mechanics and experimental quantum optics to observe interference phenomena. Thus, a number of experiments have been performed to elucidate two-photon quantum interference effects, such as the HOM effect[1,8-15] and the N00N-state interference[2,3,11,16-21]. Recently, we have reported that the two kinds of two-photon interference effects can be observed in the most generalized two-photon interferometric scheme, including a fully unfolded HOM scheme as well as a N00N-state interferometer[22].

Since the early 1990s, various apparatus for two-photon interference experiments have been utilized to investigate two-photon wavepacket interference phenomena, e.g., the Mach-Zehnder interferometer (MZI)[2,17,19,21,23,24] and the Michelson interferometer (MI)[25-27]. The majority of the experiments involving these devices were performed using two identical photons as the input state, where the two photons simultaneously entered the input port of an interferometer. However, the most interesting behavior occurs when two correlated photons are incident on the interferometer with a large time delay that is considerably longer than their coherence time[10,28,29]. Although, even in that case, if the two-photon states are still in symmetrically superposed states so as to exhibit two-photon interference effects, we can simultaneously observe the coherence properties of the single-photon and two-photon wavepackets by examining the full measured interferogram of the two-photon interference fringes. Although a number of studies have examined two-photon quantum interference experiments in a MZI or MI, further studies are required to fully elucidate the two-photon states within the interferometer arms. In addition, these studies should aim to reveal the origin of the rather complex interference fringe patterns, which contain the shapes of both the single- and two-photon wavepackets. Therefore, we aim to conduct a more comprehensive analysis on the origins of the two-photon wavepacket interference phenomena that occur when two temporally separated photons within the interferometer probabilistically generate spatially bunched/anti-bunched two-photon states.

In this paper, we report on an experimental demonstration of quantum interference effects using two kinds of two-photon state in a conventional MZI and a polarization-based MI (PMI). The two distinct two-photon states are prepared by introducing a time delay between two incident photons at the input ports of the interferometer. The two photons are well separated by a time-like interval that is longer than the coherence time of both the individual single photons and the two-photon states. Here, we consider two kinds of two-photon state within the interferometer arms in order to distinguish from the conventional HOM and N00N states, which are the *temporally separated and spatially anti-bunched* (TSSA) state and the *temporally separated and spatially bunched* (TSSB) state. The TSSA state is defined as a superposed input state with a large time delay between two single photons in two different spatial modes[10]. However, the overall state of the two photons is a symmetrically superposed state in the two spatial modes. On the other hand, the TSSB state involves two single photons with a large time delay and in the same spatial mode[21]. Experimental demonstrations employing the TSSA state have previously been performed using the polarization-entangled state[10] and, also, the frequency-entangled state[30,31]. Recently, the two-photon quantum interference of the TSSB state was successfully demonstrated revealing that the temporal separation between two sequential photons in the same spatial mode does not degrade the phase super-resolution, as in the case of the conventional N00N state[21].

**Results**

**Generation of two-photon states with temporally separated photons.** The conceptual scheme for the generation of the TSSA and TSSB two-photon states is depicted in Fig. 1. As is well known, the conventional two-photon N00N state can be easily generated via the HOM interference effect, when two identical single photons enter a balanced beamsplitter (BS) simultaneously, as shown in Fig. 1a[2,17]. In this case, the two output photons are always probabilistically bunched at one of the two spatial modes as described by

$$|1,1\rangle_{1,2} \xrightarrow{BS1} \frac{1}{\sqrt{2}}\left(|2,0\rangle_{3,4} + |0,2\rangle_{3,4}\right), \quad (1)$$

where the subscripts denote the two spatial modes of the first BS (BS1) input and output. On the other hand, the TSSA and TSSB two-photon states can be prepared by introducing a time delay, $\Delta\tau_1 = \Delta x_1/c$ where $c$ denotes the speed of light and $\Delta x_1$ is optical path-length difference between the two photons at the BS1 input stage, as shown in Fig. 1b. When the two photons are sufficiently separated from each other when relative to their coherence length, the HOM bunching effect at the BS1 output ports is no longer active. Then, the two output photons are in a state with the form

$$|1(\Delta x_1),1\rangle_{1,2} \xrightarrow{BS1} \frac{1}{\sqrt{2}}\left(|\Psi\rangle_{TSSA} + |\Psi\rangle_{TSSB}\right), \quad (2)$$

where

$$|\Psi\rangle_{TSSA} = \frac{1}{\sqrt{2}}\left[|1\rangle_3 |1(\Delta x_1)\rangle_4 - |1(\Delta x_1)\rangle_3 |1\rangle_4\right],$$

$$|\Psi\rangle_{TSSB} = \frac{i}{\sqrt{2}}\left[|1(\Delta x_1),1\rangle_3 |0\rangle_4 + \exp(i2\phi)|0\rangle_3 |1(\Delta x_1),1\rangle_4\right]. \quad (3)$$

Here, $\phi$ is the relative single-photon phase difference between the two arms of the interferometer, which can be introduced by adjusting the path-length difference $\Delta x_2$. The two kinds of two-photon state represented by Eq. (3) are probabilistically coexistent within the interferometer arms, which construct TSSA and TSSB states, respectively. Then, the final state at the second BS (BS2) output port is composed of three states, such that

$$|\Psi\rangle_{out} = \frac{1}{2}\left\{ \begin{array}{l} \sin\phi\left[|0\rangle_5 |1(\Delta x_1),1\rangle_6 - |1(\Delta x_1),1\rangle_5 |0\rangle_6\right] \\ -\cos\phi\left[|1(\Delta x_1)\rangle_5 |1\rangle_6 + |1\rangle_5 |1(\Delta x_1)\rangle_6\right] + \left[|1(\Delta x_1)\rangle_5 |1\rangle_6 - |1\rangle_5 |1(\Delta x_1)\rangle_6\right] \end{array} \right\}, \quad (4)$$

where the subscripts 5 and 6 denote the two spatial modes of the BS2 output. Here, the first term on the right-hand side represents a phase-sensitive TSSB state, while the last two terms correspond to phase-insensitive TSSA states. In the case of $\phi = 0$ or $\phi = \pi$, the output state has the form $|\Psi\rangle_{out} = |1\rangle_5 |1(\Delta x_1)\rangle_6$ or $|\Psi\rangle_{out} = |1(\Delta x_1)\rangle_5 |1\rangle_6$, respectively, which is an identical form to the BS1 input state. On the other hand, when $\phi = \pi/2$, the output state has a similar form to the BS2 input state. Although the MZI only was considered here, so as to show the output state for the input states given in Eq. (3), this result also applies to the PMI.

In this study, the observation of the two-photon interference effects obtained for the TSSA and TSSB states was performed at the MZI and PMI output ports, with $\Delta x_2$ being varied. It has been already shown that the TSSA-state interference has a phase-insensitive effect[10, 30], whereas the TSSB state can generate a resolution-enhanced phase-sensitive fringe pattern[21]. From Eq. (4), when the two input photons are injected into BS1 with a large time delay (Fig.

1b), the coincidence detection probability $P(\Delta x_2)$ at the two MZI output ports can be expressed as a superposition of the TSSA and TSSB-states interference fringes[32], such that

$$P(\Delta x_2) = N_0 \left\{ 2 + V \left[ f(\Delta x_2) + g(\Delta x_2) \cos\left(\frac{2\pi}{\lambda_p} \Delta x_2\right) \right] \right\}, \quad (5)$$

where $N_0$ is a constant, $V$ is the two-photon fringe visibility, $\lambda_p$ is the centre wavelength of the pump laser, and $f(\Delta x_2)$ and $g(\Delta x_2)$ are envelope functions corresponding to the spectral properties of the detected single- and two-photon wavepackets, respectively (see the Methods section).

**Experimental setup.** *Mach-Zehnder interferometer*: Figure 2 shows the experimental setup used to demonstrate the two-photon interference effects in an MZI. Correlated photon pairs at a telecommunication wavelength of 1.5 μm were generated through a quasi-phase-matched spontaneous parametric down-conversion (QPM-SPDC) process in a type-0 periodically-poled lithium niobate (PPLN) crystal. We used a mode-locked picosecond fiber laser (PriTel, FFL-20-HP-PRR and SHG-AF-200) as the QPM-SPDC pumping source, which had a 3.5-ps pulse duration at a 775-nm centre wavelength with a 20-MHz repetition rate. In our experiments, the average pump power was set to 20 mW. Using this setup, degenerate photon pairs were emitted with a full-opening angle of 4.6° in the noncollinear regime.

The experimental setup was composed of two fibre interferometers, and the fibre length of each MZI arm was approximately 4 m. The first fibre BS (FBS1) acted as a "state preparator" to produce the state shown in Eq. (2), while the second fibre BS (FBS2) acted as a "two-photon interferometer". The FBS2 output photons were detected after they passed through interference filters (with 6.25-nm bandwidth) via two InGaAs/InP single-photon detection modules (Id Quantique id-210), which were operated in the gated mode. Electronic trigger signals were sent from the pump to the detector gates via electric delay lines. The detector quantum efficiency and dead time were set to 15% and 10 μs, respectively. The coincidence resolving time window was set to 10 ns, which was shorter than the pulse period of 50 ns. Under these experimental conditions, the coincidence to accidental coincidence ratio was approximately 4.13. The pair production probability per pulse was obtained by dividing the accidental coincidence by the measured coincidence, which was estimated to be approximately 0.24 per pulse.

Figure 2a shows the HOM interference fringe measured at the FBS1 output port. The net visibility was found to be 99.74±2.03% from the fit of a sinc function. The fringe width was determined to be approximately 0.38 mm, which was estimated from the rectangular-shaped 6.25-nm-bandwidth interference filter. Figure 2b shows the measured two-photon interference fringe arising from the conventional N00N-state input, representative of Eq. (1) when $\Delta x_1 = 0$. The red squared symbols indicate the measured coincidence counts and are plotted as functions of $\Delta x_2$, while the gray area corresponds to the phase-sensitive oscillatory fringe pattern with 98% visibility. To observe the phase-sensitive oscillatory fringe, we measured the coincidence counts at $\Delta x_2 \approx 0$ (see inset in Fig. 2b). The measured coincidence counts are normalized and the error bars represent the Poisson statistics of the coincidence counting rates. From the sinusoidal fit to the measured data points, the net visibility is found to be 99.62±0.02%. The blue solid lines denote the envelope curves obtained from the Gaussian function having a full-width at half-maximum (FWHM) of 1.17 mm, which was determined based on the two-photon coherence length. Note that this length is dependent on the pump pulse duration and the group velocity dispersion (GVD) in the SPDC pair generation process[22].

**Experimental results.** *Two-photon interference in Mach-Zehnder interferometer with temporally separated photons*. Figure 3 shows the experimental results for the two-photon interference experiments with the two different kinds of input states shown in Fig. 1b. The conditions for the two input states before FBS2 were controlled by adjusting the first optical delay line (ODL1; $\Delta x_1$) before FBS1, and the two-photon coincidence fringes were measured for varying $\Delta x_2$ (with a 1-μm step size). When the input state was a superposition of the TSSA and TSSB states with the introduction of a large delay, $\Delta x_1 \gg x_{\text{coh.}}$, as shown in Fig. 1b, TSSA and TSSB two-photon interference fringes were observed simultaneously (Fig. 3b-e). In particular, it is worth noting that the two kinds of interference effects did not influence each other. Consequently, the two detectors (D1 and D2) probabilistically recorded the total coincidences resulting from the two kinds of interference phenomena, as expected from Eq. (5). The squared symbols represent the measured coincidence counts, which are plotted as functions of $\Delta x_2$. The gray areas correspond to the phase-sensitive oscillatory fringe patterns estimated from the cosine term in Eq. (5), while the solid lines denote the envelope curves in Eq. (5), which were determined by both the single- and two-photon spectral properties.

In Fig. 3c-e, the widths of the central fringes for simultaneous inputs of TSSA and TSSB states are equal to that of the N00N state in Fig. 3a; therefore, this width is determined by the TSSB-state interference because the coherence length of the two-photon state is much larger than those of the single-photon wavepacket in our experiment. On the other hand, the complex sinuous fringe shape is caused by the single-photon spectral property $f(\Delta x_2)$, therefore, this shape is additionally influenced by the TSSA-state interference. The dashed line in Fig. 3c represents the interference peak of the TSSA state with 49% visibility. This peak has the same width as the single-photon wavepacket, as shown in Fig. 2a, which can be obtained by randomizing the relative phase between the two interferometer arms[32,33]. Here, $f(\Delta x_2) = \text{sinc}(\Delta x_2 / \sigma_s)$, where $\sigma_s$ is related to the single-photon bandwidth, provides a measure of the TSSA-state interference fringe. Thus, $\sigma_s$ is determined by the interference filter used in the experiment only. Further, $g(\Delta x_2) = \exp\left[-\Delta x_2^2 / (2\sigma_T^2)\right]$, where $\sigma_T$ is the two-photon bandwidth, determines the size of the TSSB-state interference fringe. Therefore, the shapes and sizes of the central fringes shown in Fig. 3c-e are simultaneously determined by both the single- and two-photon coherence properties, and do not vary, even when $\Delta x_1$ is significantly longer than two-photon coherence length. When $\Delta x_2 = \pm \Delta x_1$, ordinary HOM-dip fringes are observed with 24.5% visibility, because only one-quarter of the total two-photon amplitudes contributes to the conventional HOM interference (see the side dips shown in Fig. 3b-e and the Methods section for details).

**Experimental setup.** *Polarization-based Michelson interferometer*: For the PMI, the conditions for the generation of correlated photon pairs are identical to those for the MZI. Figure 4 shows the experimental setup used to demonstrate the two-photon interference effects in a PMI. Two orthogonally polarized photons, H (horizontal polarization) and V (vertical polarization) from the SPDC source were combined using a fibre polarizing beamsplitter (FPBS) with a large delay length $\Delta x_1$; thereafter, the polarization directions of the two photons were rotated by ±45° using a half-wave plate (HWP). The PMI was composed of a polarizing beamsplitter (PBS) and two quarter-wave plates (QWPs), which had axes oriented at 45°. One of the mirrors (M2) was attached to a piezoelectric transducer (PZT) mounted on a linear translation stage, which could be used to scan the phase-sensitive interference fringe by varying $\Delta x_2$, and to extract the TSSA state from the TSSB state by randomizing the relative phase difference between the two PMI arms. To observe the two-photon interference fringe in the PMI, another PBS and a HWP with its axis oriented at 22.5° were placed at the PMI output port. Output photons from the PMI were detected by two single-photon detectors after they passed through coarse wavelength-division multiplexing (CWDM, 18-nm bandwidth) filters. In our experiment, two filter combinations (1550 and 1550 nm; 1530 and 1570 nm) were used to measure the two-photon coincidence fringes for both the degenerate and nondegenerate photon pairs. In particular, when the two-photons had different centre wavelengths, the two-photon state is in a frequency-entangled state[30,31] and it has the form of $|\Psi\rangle = 1/\sqrt{2}\left(|\omega_1\rangle_H |\omega_2\rangle_V + |\omega_2\rangle_H |\omega_1\rangle_V\right)$. Thus the TSSA state with the frequency-entangled photons and the TSSB states involving two sequentially distributed photons with different frequencies could be obtained within the PMI arms.

**Experimental results.** *Two-photon interference in polarization-based Michelson interferometer with temporally separated photons.* Figure 5 shows the two-photon interference fringes measured for two detected photons having the same wavelength of 1550 nm (a-c), and for those having different central wavelengths of 1530 and 1570 nm (d-f). When two orthogonally polarized photons with a delay of $\Delta x_1$ are injected into the PBS, as shown in Fig. 6, the two-photon states within the two PMI arms can be expressed in the same form as Eq. (3), although the polarization directions are also incorporated. For degenerate photon pairs

$$|\Psi\rangle_{\text{TSSA}} = \frac{1}{\sqrt{2}}\left[|H\rangle_T |V(\Delta x_1)\rangle_R + |V\rangle_R |H(\Delta x_1)\rangle_T\right],$$
$$|\Psi\rangle_{\text{TSSB}} = \frac{1}{\sqrt{2}}\left[|H\rangle_T |H(\Delta x_1)\rangle_T + \exp(i2\phi)|V\rangle_R |V(\Delta x_1)\rangle_R\right], \quad (6)$$

and for nondegenerate photon pairs

$$|\Psi\rangle_{\text{TSSA}} = \frac{1}{2}\begin{bmatrix} |\omega_1,H\rangle_T |\omega_2,V(\Delta x_1)\rangle_R + |\omega_1,V\rangle_R |\omega_2,H(\Delta x_1)\rangle_T \\ +|\omega_2,H\rangle_T |\omega_1,V(\Delta x_1)\rangle_R + |\omega_2,V\rangle_R |\omega_1,H(\Delta x_1)\rangle_T \end{bmatrix},$$

$$|\Psi\rangle_{\text{TSSB}} = \frac{1}{2}\begin{bmatrix} |\omega_1,H\rangle_T |\omega_2,H(\Delta x_1)\rangle_T + \exp(i2\phi)|\omega_1,V\rangle_R |\omega_2,V(\Delta x_1)\rangle_R \\ +|\omega_2,H\rangle_T |\omega_1,H(\Delta x_1)\rangle_T + \exp(i2\phi)|\omega_2,V\rangle_R |\omega_1,V(\Delta x_1)\rangle_R \end{bmatrix},$$ (7)

where the subscripts $T$ (transmission) and $R$ (reflection) denote the two spatial modes of the PBS output and $\omega_{1,2}$ represent the central frequencies of the two correlated photons. Here, it is worth noting that the two-photon amplitudes in Eq. (7), i.e., $|\omega_2,H\rangle_T |\omega_1,H(\Delta x_1)\rangle_T$ and $|\omega_2,V\rangle_R |\omega_1,V(\Delta x_1)\rangle_R$ for the TSSB state and $|\omega_2,H\rangle_T |\omega_1,V(\Delta x_1)\rangle_R$ and $|\omega_2,V\rangle_R |\omega_1,H(\Delta x_1)\rangle_T$ for the TSSA state, can be ignored when the two input photons are well separated from each other compared with the coincidence resolving time window $T_R$ ($\Delta x_1/c \gg T_R$). Similar to the MZI scenario, these two kinds of two-photon states are probabilistically coexistent, whether the two photons are in the same spatial mode (TSSB) and have the same polarization or if they are in two different spatial modes (TSSA) and are orthogonally polarized.

In the experiment, we set the $\Delta x_1$ of the vertically polarized photons to 3.2 mm. Figure 5a,d show the two-photon interference fringes measured by varying $\Delta x_2$ with 1-μm step size. The square symbols are the measured coincidence counts and are plotted as functions of $\Delta x_2$, while the gray are correspond to the phase-sensitive oscillatory fringe patterns. The solid lines denote the envelope curves obtained from Eq. (5). To observe the phase super-resolved fringe, we scanned one of the mirrors (M2) using the PZT actuator, for $\Delta x_2 \approx 0$. Figure 5b,e show the two-photon interference fringes with 26.21±0.58% (30.97±1.41%) visibility measured for degenerate (nondegenerate) photon pairs. If we randomized the relative phase between the two interferometer arms, we could then extract the interference fringe of the TSSA state from that of the TSSB state; this is because the TSSB state is very sensitive to the relative phase difference, while the TSSA state has no phase-sensitive interference fringe[32,33]. Figure 5c,f show the measured TSSA two-photon interference fringes as functions of $\Delta x_2$ that were obtained when a DC voltage value of 0~90 V was applied to the PZT actuator with a frequency of 10 Hz. The fringe visibility was found to be 39.11±3.38% (43.82±2.03). In the experiment, slightly lower visibility is mainly due to the imperfect alignment, instability in the interferometer including PZT actuator, and imperfection of the polarization optical components such as the PBS and wave plates.

## Discussion

Although various kinds of two-photon interference experiments involving correlated photons and using specific interferometers have been performed over the past three decades, the interference phenomena considered in this work, which arise from two-photon states composed of temporally separated photons, have not been fully analyzed. Moreover, a two-photon interference experiment with temporally separated photons in a Michelson interferometer has not yet been reported. Note that studies related to our MZI experiment have been conducted previously, using a coherently recurrent pump mode as the SPDC pump[34,35]. Those studies have presented some mathematical analyses of the interference phenomena. However, the present study provides fully generalized mathematical analyses (in Methods section), and qualitative and more comprehensive explanations of both the interference fringe shapes in the case of simultaneous inputs of the TSSA and TSSB states, and of the phase-insensitive side peak fringes arising from ordinary HOM-state inputs. In addition, the dispersion cancellation effects observed in two-photon interference experiments involving frequency-anticorrelated photon pairs in a HOM scheme can also be explained in the context of TSSA state interference, as depicted in Fig. 1b[32,36]. Indeed, many of the two-photon quantum interference phenomena can be more clearly understood by employing probabilistically coexisting states such as the TSSA and TSSB two-photon states.

In conclusion, we have demonstrated two-photon interference experiments involving temporally separated photons in an MZI and a PMI. We have introduced the concept of TSSA and TSSB two-photon states in order to distinguish from the conventional two-photon state with no time delay between the two constituent photons. By introducing a large time delay in the input stage of the interferometer, we successfully prepared two kinds of symmetrically superposed states of TSSA and TSSB two-photon states. The two-photon interference fringes measured for the two different kinds of two-photon states revealed the interferometric properties of both the single- and the two-photon wavepackets simultaneously, within a single interferometric setup. Further experimental investigation and related analysis can further clarify the origin of the two-photon interference effects in both the MZI and PMI. We believe that the present results will enable a more comprehensive understanding of the interference phenomena involving correlated quantum particles.

## Methods

**Theoretical description.** The quantum state of the photon pair source in Fig. 1 can be descripted as $|\Psi\rangle = \iint d\omega_1 d\omega_2 \Phi(\omega_1, \omega_2) \hat{a}_1^\dagger(\omega_1) \hat{a}_2^\dagger(\omega_2) |0\rangle$, where $\Phi(\omega_1, \omega_2)$ denotes the two-photon wave function, $\hat{a}_i^\dagger(\omega_j)$ is the creation operator of frequency $\omega_i$ at path $j$, and $|0\rangle$ is vacuum state. The two-photon coincidence counting rate between path 5 and 6 in Fig. 1 is proportional to the time-averaged value of the photon detection probability defined as

$P_{5,6}(t_1,t_2) = |\langle 0|\hat{E}_5^{(+)}(t_1)\hat{E}_6^{(+)}(t_2)|\Psi\rangle|^2$, where $\hat{E}_k^{(+)}(t_l)$ denotes the positive part of the electric field operator at time $t_l$ in path $k$. The electric field operators for paths 5 and 6 are superposed of them for path 1 and 2 in Fig. 1. If we assume that paths 2 and 4 have optical delay lines denoted $\tau_1$ and $\tau_2$, respectively, the normalized coincidence counting probability is calculated as

$$P = \frac{1}{2} + \frac{1}{8}Re\left\{\begin{array}{l} 2\iint \Phi^*(\omega_1,\omega_2)\Phi(\omega_1,\omega_2)\left[e^{i\tau_2(-\omega_1+\omega_2)} + e^{i\tau_2(\omega_1+\omega_2)}\right]d\omega_1 d\omega_2 \\ +\iint \Phi^*(\omega_2,\omega_1)\Phi(\omega_1,\omega_2)\left[2e^{i\tau_1(\omega_1-\omega_2)+i\tau_2(\omega_1+\omega_2)} - e^{i(\tau_1+\tau_2)(\omega_1-\omega_2)} - e^{i(\tau_1-\tau_2)(\omega_1-\omega_2)}\right]d\omega_1 d\omega_2\end{array}\right\}. \quad (8)$$

For a comprehensive understanding, we assume that the two-photon wave function is symmetric, $\Phi(\omega_1,\omega_2) = \Phi(\omega_2,\omega_1)$, and the time delay $\tau_1$ is zero or much larger than two-photon coherence time $\tau_{coh.}$. When $\tau_1$ is zero, Eq. (8) can be simplified as

$$P = \frac{1}{2}\left\{1 + Re\left[\iint |\Phi(\omega_1,\omega_2)|^2 e^{i\tau_2(\omega_1+\omega_2)}d\omega_1 d\omega_2\right]\right\}. \quad (9)$$

It represents phase sensitive two-photon interference as shown in Fig. 3a[37]. The envelope function of the interference patterns is related with Fourier transform of two-photon wave function for the direction of $\omega_1 = \omega_2$, so that the width of the interference pattern is decided by the spectral bandwidth of photon pair source[22]. If $\tau_1$ is much larger than $\tau_{coh.}$, the interference patterns are revealed only when $\tau_2$ is around 0 or $\pm\tau_1$. For the region of $|\tau_2 = \Delta\tau| < \tau_{coh.}$, Eq. (8) is simplified as

$$P = \frac{1}{2}\left\{1 + \frac{1}{2}Re\left[\iint |\Phi(\omega_1,\omega_2)|^2 \left(e^{i\Delta\tau(\omega_1-\omega_2)} + e^{i\Delta\tau(\omega_1+\omega_2)}\right)d\omega_1 d\omega_2\right]\right\}, \quad (10)$$

which includes phase insensitive HOM peak and phase sensitive two-photon interferences due to the terms of $(\omega_1 - \omega_2)$ and $(\omega_1 + \omega_2)$, respectively, and their amplitudes are reduced to a half. The width of HOM interference is related with the single-photon coherence time and the Fourier transform of two-photon wave function for the direction of $\omega_1 = -\omega_2$. It is decided by phase matching condition of the SPDC process and filter bandwidth used in the experiment. For the region of $|\tau_2 \pm \tau_1 = \Delta \tau| < \tau_{coh.}$, Eq. (8) is simplified as

$$P = \frac{1}{2}\left\{1 - \frac{1}{4} Re\left[\iint |\Phi(\omega_1, \omega_2)|^2 e^{-i\Delta\tau(\omega_1 - \omega_2)} d\omega_1 d\omega_2\right]\right\}. \quad (11)$$

It only shows HOM dip interference with quarter amplitude. Eqs. (9-11) clearly describe the experimental results in Fig. 3, and show the definitions of envelope functions $f(\Delta x_2)$ and $g(\Delta x_2)$ in Eq. (5). We note that Eq. (8) is not simplified and shows complicated interference patterns under a condition of $0 < \tau_1 < \tau_{coh.}$. For the case of non-degenerated photon pairs (frequency-entangled states), the assumption of the symmetry, $\Phi(\omega_1, \omega_2) = \Phi(\omega_2, \omega_1)$ is still satisfied. The only difference from degenerated case is that the beating fringes also arise in HOM interference due to the wavelength difference between two photons, as shown in Fig. 5f.

**Figure 1. Generation of two-photon states with temporally separated photons.** Conceptual scheme for generation of (**a**) conventional N00N state from HOM interference effect at BS1 output port and (**b**) superposed state of TSSA and TSSB states with large time delay between two photons at BS1 input port.

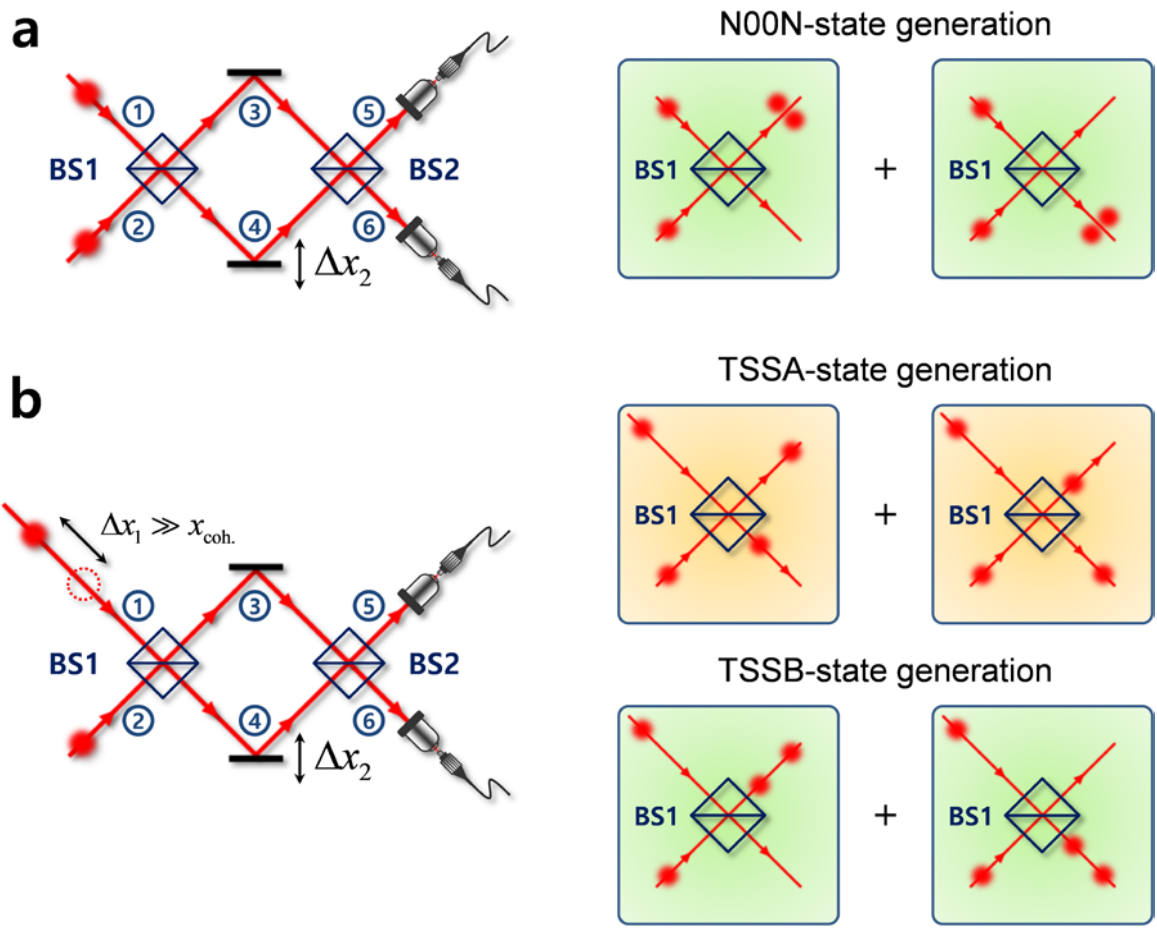

**Figure 2. Experimental setup.** *Mach-Zehnder interferometer*: Pump: picosecond mode-locked fibre laser (3.5 ps, 20 MHz, 775 nm, 20 mW); PBS: polarizing beamsplitter; L1, L2: spherical lenses with 200-mm focal length; PPLN: periodically-poled lithium niobate crystal (length 10 mm, grating period 19.2 μm, temperature 40°C); DM: dichroic mirror (T1550 nm/R775 nm); L3: aspherical lens with 8-mm focal length; PC: polarization controller; ODL: optical delay line; FBS: fibre beamsplitter 50/50; IF: interference filter with 6.25 nm bandwidth, D1, D2: gated-mode single-photon detection modules (Id Quantique id-210). (**a**) HOM-dip fringe measured at FBS1 output port as a function of delay length $\Delta x_1$. (**b**) N00N-state fringe measured at FBS2 output port as a function of path-length difference $\Delta x_2$. The inset shows the measured interference fringe at $\Delta x_2 \approx 0$.

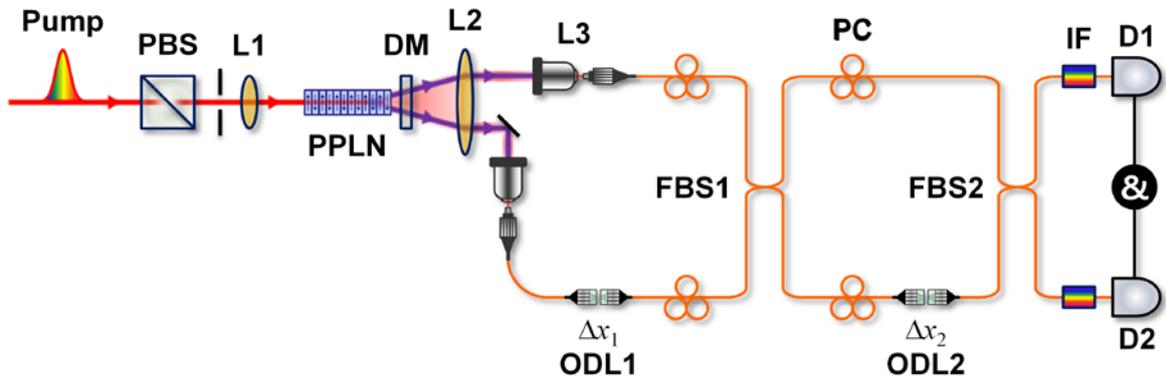

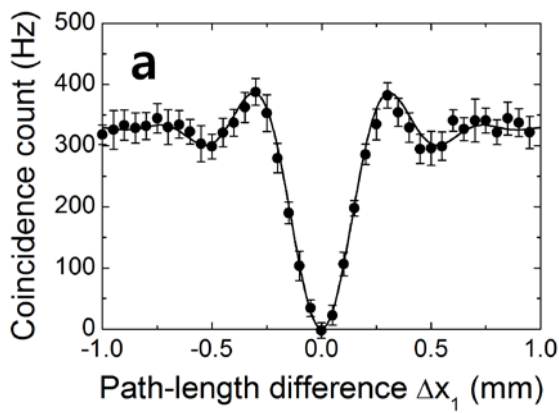
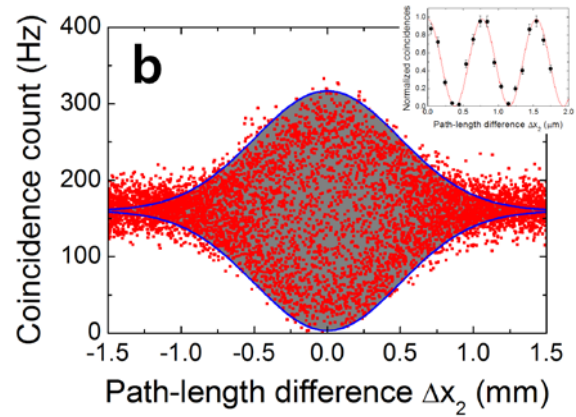

**Figure 3. Two-photon interference fringes measured in Mach-Zehnder interferometer.**
Interference fringes obtained for (**a**) conventional N00N-state input ($\Delta x_1 = 0$) and (**b**-**e**) simultaneous inputs of TSSA and TSSB states for various $\Delta x_1$ delay positions. The side dips represent ordinary HOM interference fringes obtained for path-length difference $\Delta x_2 = \pm \Delta x_1$.

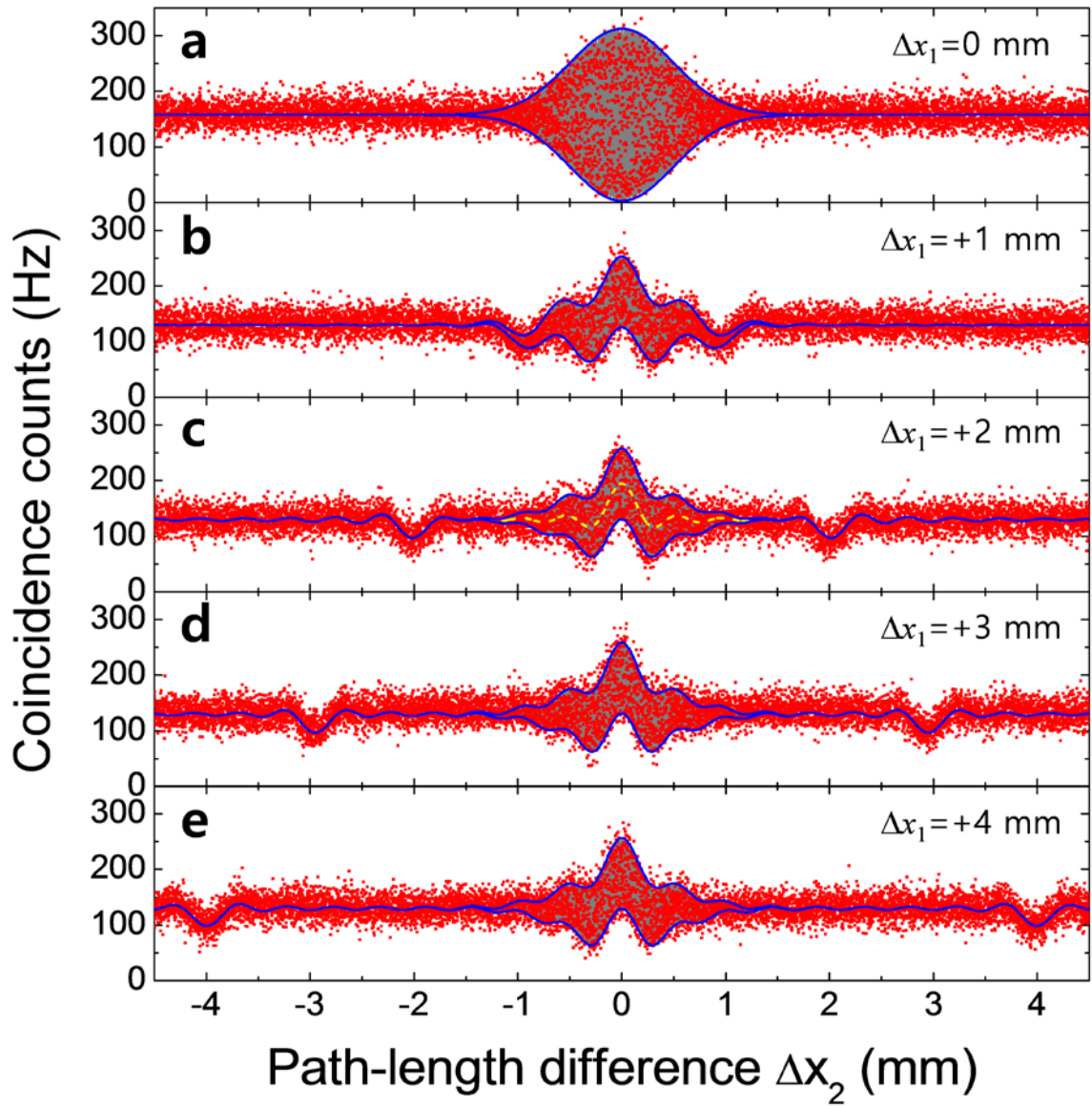

**Figure 4. Experimental setup.** *Polarization-based Michelson interferometer*: Two orthogonally polarized photons were combined using a fibre polarizing beamsplitter (FPBS) with a delay length $\Delta x_1$. HWP: half-wave plate; PBS: polarizing beamsplitter; QWP: quarter-wave plate; M: mirror; FC: single-mode fibre coupler. A piezoelectric transducer (PZT) was used to scan the phase-sensitive interference fringe and to randomize the relative phase between two interferometer arms.

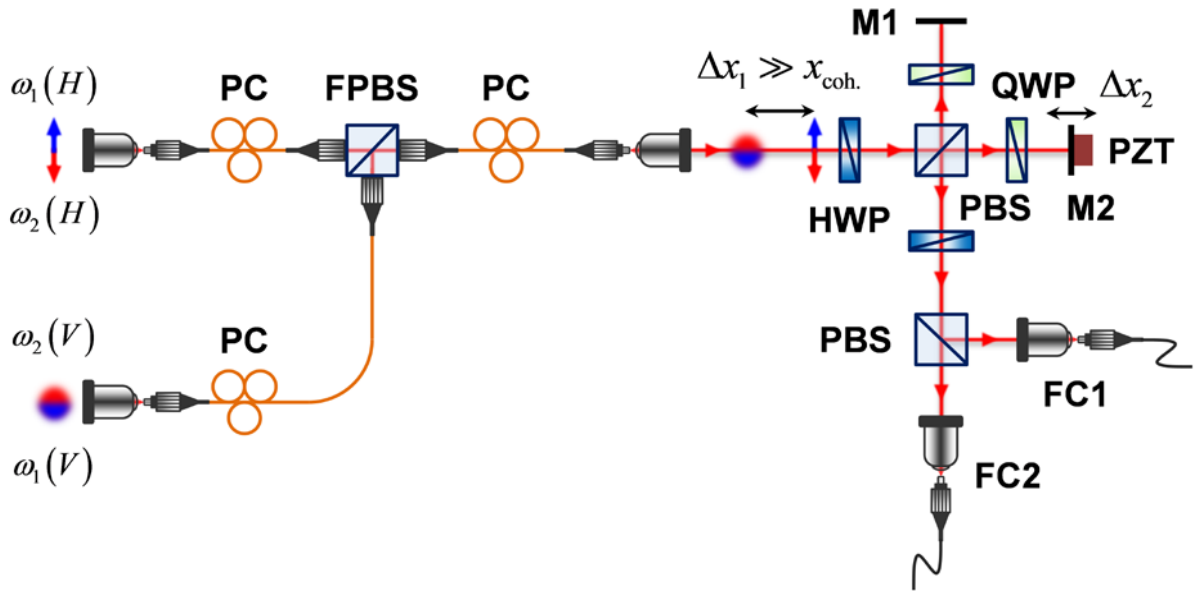

**Figure 5. Two-photon interference fringes measured in polarization-based Michelson interferometer.** Interference fringes obtained for simultaneous inputs of TSSA and TSSB states with (**a**-**c**) degenerate photon pairs and (**d**-**f**) nondegenerate photon pairs.

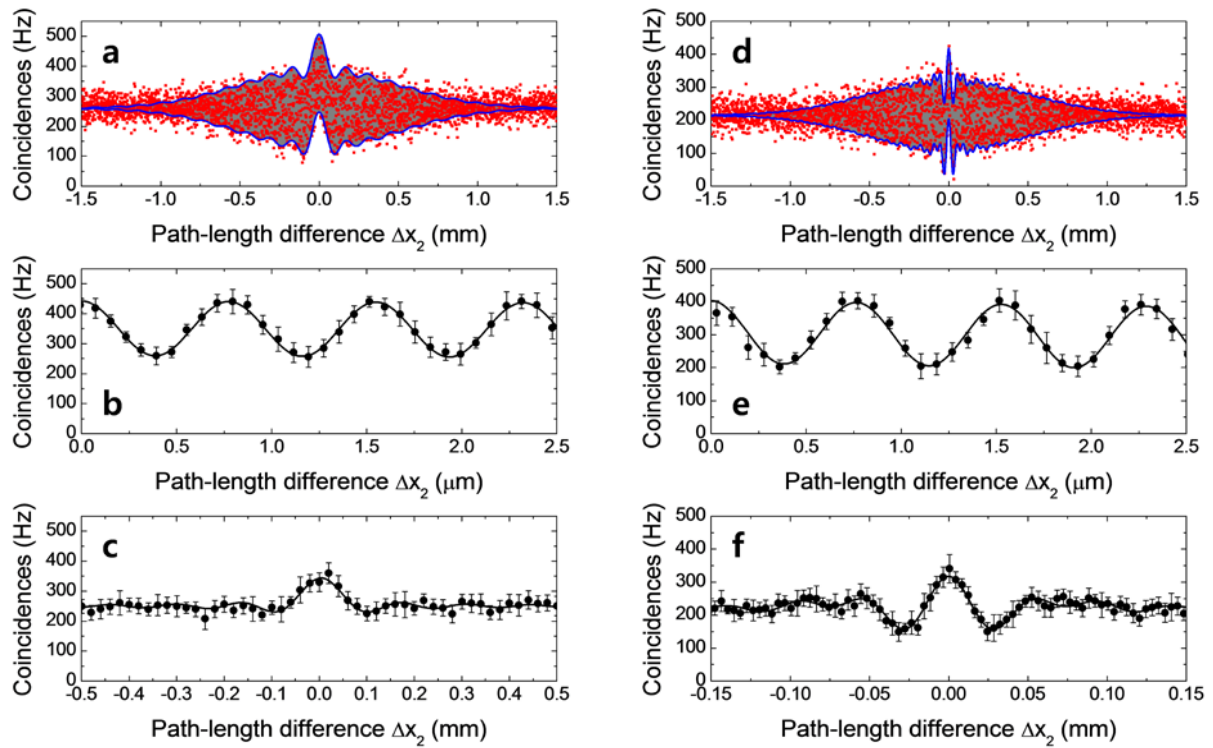

**Figure 6. Generation of two-photon states with temporally separated photons in polarization-based Michelson interferometer.** For (**a,b**) degenerate photon pairs and (**c,d**) nondegenerate photon pairs. In the case of the TSSA state, the two photons are in two different spatial modes. For the TSSB state, the two photons are in the same spatial mode.

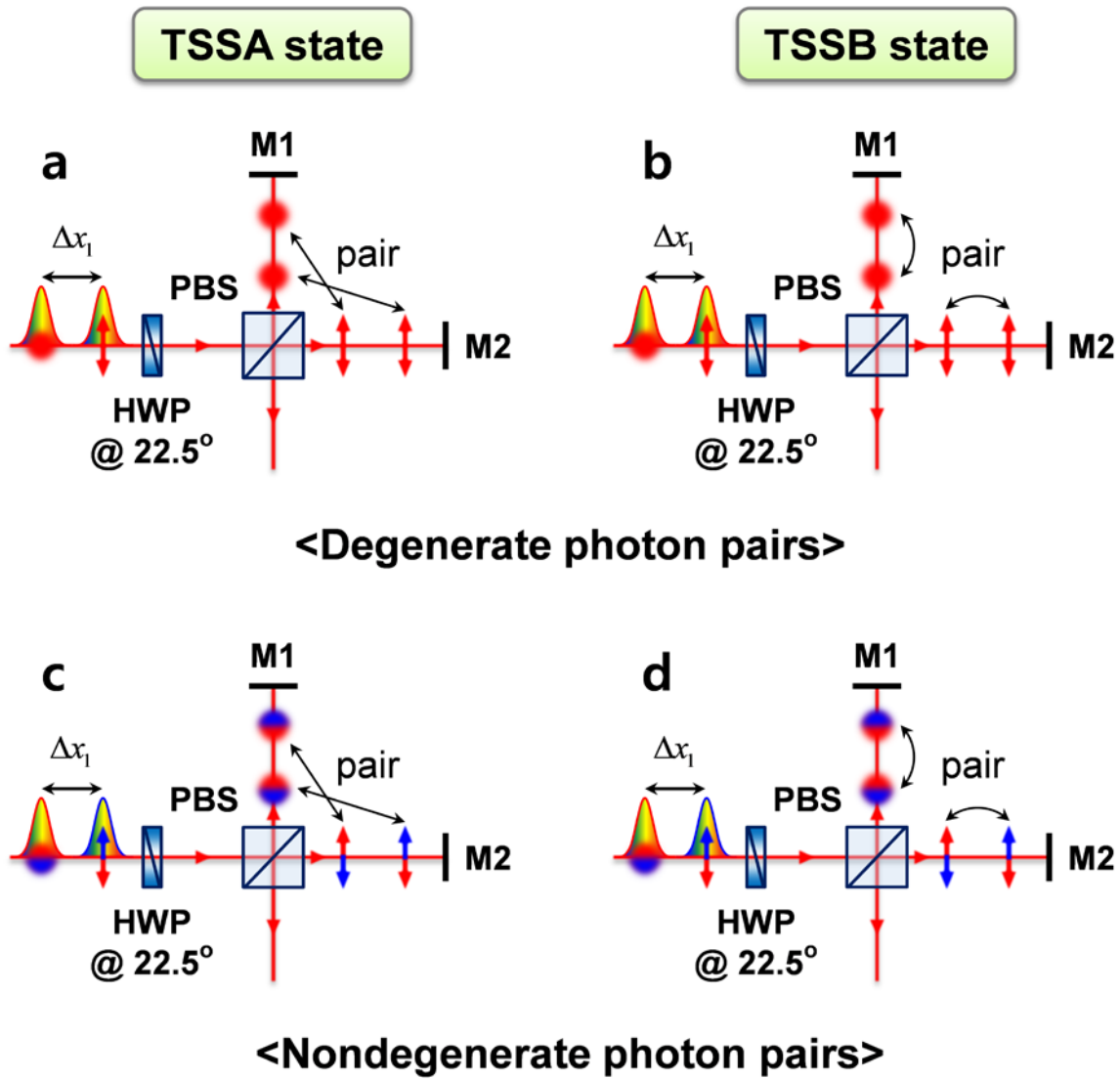